\documentclass{article}
\usepackage{amsmath,amssymb,latexsym,graphicx}

\begin{document}

\begin{flushright}
	KUNS-1967 \\
	YITP-05-16 \\
	OIQP-05-04 \\
	hep-th/0504173
\end{flushright}

\vspace{0.5cm}

\begin{center}
{\Large\bf Hole Theory of Boson Sea}\\
\vspace{1.5cm}
{\large Yoshinobu H}ABARA\\
\vspace{0.2cm}
{\it  Department of Physics, Graduate School of Science,}\\
{\it  Kyoto University, Kyoto 606-8502, Japan}\\
\vspace{1cm}
{\large Holger B. N}IELSEN\\
\vspace{0.2cm}
{\it  Niels Bohr Institute, University of Copenhagen,}\\
{\it  17 Blegdamsvej Copenhagen \o, Denmark}\\
\vspace{0.3cm}
and\\
\vspace{0.3cm}
{\large Masao N}INOMIYA\footnotemark \\
\vspace{0.2cm}
{\it  Yukawa Institute for Theoretical Physics,}\\
{\it  Kyoto University, Kyoto 606-8502, Japan}
\end{center}

\footnotetext{Also working at Okayama Institute for Quantum Physics, Kyoyama-cho 1-9, Okayama-city 700-0015, Japan}

\vspace{0.5cm}

\begin{abstract}
We propose a hole theory for bosons, in which, analogous to fermions, a hole produced by the annihilation of one negative energy boson is an anti-particle. We show that the boson vacuum indeed also consists of a sea in which all negative energy states are filled and the density of probability of the Klein-Gordon theory is positive definite also for the negative energy solution. This formalism is obtained by introducing the notion of a double harmonic oscillator, which is constructed by extending the condition imposed on the wave function. This double harmonic oscillator contains not only positive energy states but also negative energy ones. The physical result obtained from our method is consistent with that of the ordinary second quantization formalism. Our formulation is also consistent with the supersymmetric point of view. We finally suggest applications of our method to the anomalies of boson theories and the string theories.
\end{abstract}

\newpage

\section{Introduction}

\vspace{0.5cm}

One of the long-standing problems in field theories is to clarify a discrepancy of procedure of quantizing bosons and fermions. Both theories should satisfy the energy-momentum relation of A. Einstein, 

\begin{align*}
	E^2-\vec{p}^{\> 2}=m^2 \> \Longleftrightarrow \> 
	E=\pm \sqrt{\vec{p}^{\> 2}+m^2}.
\end{align*}

\noindent One implication of this relation is that we have not only positive energy states but also negative energy ones. It is not easy to obtain a physical understanding of negative energy states and the implications of their existence. In fact, the existence of particle states with negative energy indicates bottomlessness in energy. For instance, a particle can emit another particle, such as a photon etc., and thereby decay into a lower energy state. It is thus seen that the particle in question cannot exist as a stable state. This has been regarded as one of the fundamental problems in field theories and has vexed elementary particle physicists for many years (for a historical account, see Ref.~\cite{weinberg}). However in fermion theory, in 1930 P. A. M. Dirac succeeded in obtaining a physically consistent particle picture. His well-known interpretation is the following: The fermion vacuum is not an empty state. Rather, all the negative energy states are filled, forming what is called the ``Dirac sea", and thus it is impossible for a positive energy particle to fall into the negative energy states, due to Pauli's exclusion principle. Furthermore, if one negative energy particle disappears, and a hole is thereby created in the Dirac sea, this hole is interpreted as a positive energy particle relative to the surrounding Dirac sea. This hole therefore can be considered a positive energy anti-particle. This surprising prediction was confirmed by the discovery of the positron, the anti-particle of the electron. In this manner, the problem of negative energy states for fermions was solved. Indeed, the chiral anomaly, which is a peculiar quantum effect in fermion theories, is understood and re-derived as one of the physical phenomena resulting from the Dirac sea~\cite{nn}.

On the other hand, it seems that, in the ordinary understanding of present day particle physics, we are able to obtain consistent boson field theories by re-interpreting creation and annihilation of a negative energy particle as the annihilation and creation of a positive energy particle. The negative energy vacuum in such a theory is empty, like the vacuum of a positive energy solution.

However we believe that the ordinary boson vacuum theory described above consists basically of a re-reading of the creation and annihilation operators, which are essential tools in quantum theory. This re-reading is nothing but a re-naming of the operators mathematically, and hence we conclude that the ordinary boson theory does not provide a physically satisfactory explanation of the boson vacuum structure. Moreover, in view of supersymmetry, which plays an essential role in recent elementary particle theories, such as superstring theories and supersymmetric grand unified theories, we feel that there is a problem with regard to the fundamental difference between the fermion vacuum and the boson one in the standard treatment: In the fermion vacuum, the negative energy states are totally filled, while the boson vacuum is empty.

It is the purpose of the present article to propose a method about how to construct a ``boson sea"\footnotemark analogous to the Dirac sea by investigating negative energy solutions in detail. Regrettably, in our previously published papers~\cite{ynn,nn3}, the Hilbert space spanned by solutions with the negative energy has a negative norm so that it is impossible to provide physical particle picture out of these solutions. In the present paper we completely solved the negative norm problem to succeed to construct the Hilbert spaces (with the positive norm by its definition) of the positive and negative energy solutions. In our new method as the true vacuum of the boson theory the negative energy sea is completely filled. In the course of this study, supersymmetry plays an important role. However, we generalize the formalism so that our proposed theory holds in general, even in non-supersymmetric boson theories. 

\footnotetext{The idea of a boson sea was proposed by two of the present authors (H.B.N. and M.N.) a few years ago~\cite{nn2}. However, in Ref.~\cite{nn2}, there is used a slightly different notation from the present article with respect to enumerating the extrapolated on negative excitation number states. So $|-n\rangle$ is called $|-n+1\rangle$ in the present article. Then we have two states $|0_+\rangle$ and $|0_-\rangle$ in the present paper with the name $0$. Of course $|0_-\rangle =|-1\rangle \Big|_{\text{Ref.~\cite{nn2}}}$ and $|0_+\rangle =|0\rangle \Big|_{\text{Ref.~\cite{nn2}}}$.}

\vspace{0.5cm}

In Section 2, in view of supersymmetry, which asserts the equivalence of bosons and fermions in a certain sense, we investigate the boson vacuum that should be supersymmetric to the fermion vacuum, i.e. the Dirac sea. As a concrete example, we consider the theory of an $N=2$ matter multiplet called a hypermultiplet. In this theory, the condition for the boson vacuum is that it vanishes through the creation of a negative energy particle. This may seem counterintuitive, but in fact this condition is equivalent to that for the fermion vacuum. In Section 3, we introduce a double harmonic oscillator in the first quantized theory. In the ordinary theory, there appear only states with positive energy. However, we extend the condition for the wave function and in so doing obtain negative energy states as well as positive energy ones. It is argued that these new negative energy states are precisely what is needed to describe bosons with negative energy. We then show that the boson vacuum is also a kind of filled state, analogous to the fermion vacuum, with the negative energy states filled by bosons. In Section 4, we apply the double harmonic oscillator to the second quantized boson system. We then show that true boson vacuum is obtained from a vacuum analogous to the empty Dirac sea which we associate with a sector by an operation (67) to be seen below. We discuss a new particle picture in which the hole produced by annihilating a negative energy particle is an anti-particle that is observable. We confirm that our theory is physically consistent by considering the energies of states. In Section 5, we discuss the negative energy solution of the Klein-Gordon equation, i.e. one particle state with negative energy. It is well-known that this one particle state has the indefinite norm, in other words, the negative density of probability. We show that this difficulty can be resolved due to our method. Section 6 is devoted to a conclusion and a brief overview of possible future developments.

\section{Boson vacuum in a supersymmetric theory}

\vspace{0.5cm}

In the present section, we consider an ideal world in which supersymmetry holds exactly. Then, it is natural to believe that in analogy to the true fermion vacuum the true boson vacuum is a state in which all negative energy states are occupied. To investigate the details of the vacuum structure of bosons needed to realize supersymmetry with the Dirac sea, we utilize the $N=2$ matter multiplet called a hypermultiplet~\cite{sohnius,west}. In fact, we construct the Noether current from the supersymmetric action, and by requiring that the entire system be supersymmetric, we derive the properties of the boson vacuum, while the fermion vacuum is taken to be the Dirac sea.

Hereafter, the Greek indices $\mu ,\nu ,\cdots$ are understood to run from 0 to 3, corresponding to the Minkowski space, and the metric is given by $\eta^{\mu \nu}=diag(+1,-1,-1,-1)$.

\subsection{$N=2$ matter multiplet: Hypermultiplet}

\vspace{0.5cm}

Let us summarize the necessary part of $N=2$ supersymmetric field theory in the free case, in order to reveal the difficulty of making supersymmetric description analogous to a not yet filled Dirac sea. The hypermultiplet is the simplest multiplet that is supersymmetric and involves a Dirac fermion. It is written 

\begin{align}
	\phi =(A_1, A_2;\> \psi ;\> F_1, F_2), 
\end{align}

\noindent where $A_i$ and $F_i\> (i=1,2)$ denote complex scalar fields, and the Dirac field is given by $\psi$. The multiplet (1) transforms under a supersymmetric transformation as 

\begin{align}
	& \delta A_i=2\bar{\xi}_i\psi , \nonumber \\
	& \delta \psi =-i\xi_iF_i-i\gamma^{\mu}\partial_{\mu}\xi_iA_i, 
	\nonumber \\
	& \delta F_i=2\bar{\xi_i}\gamma^{\mu}\partial_{\mu}\psi ,
\end{align}

\noindent where $\gamma^{\mu}$ denotes the four-dimensional gamma matrices, with $\{ \gamma^{\mu},\gamma^{\nu}\} =2\eta^{\mu \nu}$. The commutator of the supersymmetry transformations is given by 

\begin{align}
	[\delta^{(1)},\delta^{(2)}]=2i\bar{\xi}_i^{(1)}\gamma^{\mu}
	\xi_i^{(2)}\partial_{\mu}+2i\bar{\xi}_i^{(1)}\xi_i^{(2)}\delta_Z,
\end{align}

\noindent where $\delta^{(1)}$ and $\delta^{(2)}$ are the supersymmetry transformations associated with $\xi^{(1)}$ and $\xi^{(2)}$ respectively, and $\delta_Z$ represents the $N=2$ supersymmetry transformation with central charge $Z$. Its explicit transformation law takes the form 

\begin{align}
	\left\{ \begin{array}{l}
	\delta_ZA_i=F_i, \\
	\delta_Z\psi =\gamma^{\mu}\partial_{\mu}\psi , \\
	\delta_ZF_i=-\partial_{\mu}\partial^{\mu}A_i.
	\end{array} \right.
\end{align}

\noindent This transformation law satisfies the condition 

\begin{align}
	\delta_Z^2=-\partial_{\mu}\partial^{\mu}=P_{\mu}P^{\mu}=m^2,
\end{align}

\noindent where $m$ denotes the multiplet mass.

A supersymmetric scalar density can be constructed from two hypermultiplets 

\begin{align*}
	& \phi =(A_1, A_2;\> \psi ;\> F_1, F_2), \\
	& \rho =(B_1, B_2;\> \chi ;\> G_1, G_2)
\end{align*}

\noindent as the inner product 

\begin{align*}
	(\bar{\rho}\cdot \phi )\equiv iB_i^{\dagger}F_i-iG_i^{\dagger}A_i
	+2\bar{\chi}\psi ,
\end{align*}

\noindent up to total derivatives. From this scalar density, we obtain the Lagrangian density of the hypermultiplet, 

\begin{align}
	\mathcal{L} & =\frac{i}{2}(\bar{\phi}\cdot \delta_{Z}\phi)
	+\frac{m}{2}(\bar{\phi}\cdot \phi) \nonumber \\
	& =\frac{1}{2}\partial_{\mu}A_i^{\dagger}\partial^{\mu}A_i
	+\frac{1}{2}F_i^{\dagger}F_i+i\bar{\psi}\gamma^{\mu}\partial_{\mu}\psi
	+m\left[\frac{i}{2}A_i^{\dagger}F_i
	-\frac{i}{2}F_i^{\dagger}A_i+\bar{\psi}\psi \right].
\end{align}

\noindent The hermitean form of (6) is given by 

\begin{align}
	\mathcal{L} & =\frac{1}{2}\partial_{\mu}A_i^{\dagger}\partial^{\mu}
	A_i+\frac{1}{2}F_i^{\dagger}F_i+\frac{i}{2}\bar{\psi}\gamma^{\mu}
	\partial_{\mu}\psi -\frac{i}{2}\partial_{\mu}\bar{\psi}\gamma^{\mu}
	\psi +m\left[\frac{i}{2}A_i^{\dagger}
	F_i-\frac{i}{2}F_i^{\dagger}A_i+\bar{\psi}\psi \right].
\end{align}

To derive the Noether currents whose charges generate the supersymmetry transformation, we consider a variation under the supersymmetry transformation (2), 

\begin{align}
	\delta \mathcal{L}=\bar{\xi}_i\partial_{\mu}K_i^{\mu}
	+\partial_{\mu}\bar{K}_i^{\mu}\xi_i,
\end{align}

\noindent where $K_i^{\mu}$ is given by 

\begin{align}
	K_i^{\mu}\equiv \frac{1}{2}(\gamma^{\mu}\gamma^{\nu}\psi 
	\partial_{\nu}A_i^{\dagger}+im\gamma^{\mu}\psi A_i^{\dagger}).
\end{align}

\noindent Thus, the Noether current $J_i^{\mu}$ is written 

\begin{align}
	& \bar{\xi}_iJ_i^{\mu}+\bar{J}_i^{\mu}\xi_i=
	\frac{\delta L}{\delta (\partial_{\mu}\phi)}\delta \phi 
	-\left(\bar{\xi}_iK_i^{\mu}+\bar{K}_i^{\mu}\xi_i\right) \nonumber \\
	& \qquad \qquad \quad =\bar{\xi}_i
	\big(\gamma^{\nu}\gamma^{\mu}\psi \partial_{\nu}
	A_i^{\dagger}-im\gamma^{\mu}\psi A_i^{\dagger}\big)+\big(\partial_{\nu}
	A_i\bar{\psi}\gamma^{\mu}\gamma^{\nu}+imA_i\bar{\psi}\gamma^{\mu}\big)
	\xi_i, \\
	& \quad \Longrightarrow J_i^{\mu}=\gamma^{\nu}\gamma^{\mu}\psi 
	\partial_{\nu}A_i^{\dagger}-im\gamma^{\mu}\psi A_i^{\dagger}.
\end{align}

We could attempt to think of treating the bosons analogous to the fermions by imagining that the creation operators $a^{\dagger}(\vec{k})$ of the anti-bosons were really annihilation operators in some other formulation, but, as we shall see, such an attempt leads to some difficulties. If we could indeed do so, analogously to the pre-Dirac sea filling fermion field consisting solely of annihilation operators, we would write also a boson field in terms of only annihilation operators formally as follows 

\begin{align}
	& A_i(x)=\int \frac{d^3\vec{k}}{\sqrt{(2\pi )^32k_0}}\left\{
	a_{i+}(\vec{k})e^{-ikx}+a_{i-}(\vec{k})e^{ikx}\right\}, \\
	& \psi(x)=\int \frac{d^3\vec{k}}{\sqrt{(2\pi )^32k_0}}\sum_{s=\pm}
	\left\{b(\vec{k},s)u(\vec{k},s)e^{-ikx}+d(\vec{k},s)v(\vec{k},s)e^{ikx}
	\right\}.
\end{align}

\noindent Here, $k_0\equiv \sqrt{\vec{k}^2+m^2}$ is the energy of the particle, and $s\equiv \frac{\vec{\sigma}\cdot \vec{k}}{|\vec{k}|}$ denotes the helicity. Particles with positive and negative energy are described by $a_{i+}(\vec{k}), b(\vec{k},s)$ and $a_{i-}(\vec{k}), d(\vec{k},s)$, respectively. The commutation relations between these field modes are derived as 

\begin{align}
	& \left[a_{i+}(\vec{k}),a_{j+}^{\dagger}(\vec{k}^{\prime})\right]=
	+\delta_{ij}\delta^3(\vec{k}-\vec{k}^{\prime}), \\
	& \left[a_{i-}(\vec{k}),a_{j-}^{\dagger}(\vec{k}^{\prime})\right]=
	-\delta_{ij}\delta^3(\vec{k}-\vec{k}^{\prime}), \\
	& \left\{b(\vec{k},s),b^{\dagger}(\vec{k}^{\prime},s^{\prime})\right\}
	=+\delta_{ss^{\prime}}\delta^3(\vec{k}-\vec{k}^{\prime}), \\
	& \left\{d(\vec{k},s),d^{\dagger}(\vec{k}^{\prime},s^{\prime})\right\}
	=+\delta_{ss^{\prime}}\delta^3(\vec{k}-\vec{k}^{\prime}), 
\end{align}

\noindent with all other pairs commuting or anti-commuting. Note that the right-hand side of the commutation relation (15), for negative energy bosons, has the opposite sign of (14), for positive energy bosons. In the standard context, these creation and annihilation operators have the following interpretations: 

\begin{align*}
	& \left\{ \begin{array}{l}
	a_{i+}(\vec{k})\text{ annihilates a positive energy boson,} \\
	a_{i+}^{\dagger}(\vec{k})\text{ creates a positive energy boson,} \\
	a_{i-}(\vec{k})\text{ annihilates a negative energy boson,} \\
	a_{i-}^{\dagger}(\vec{k})\text{ creates a negative energy boson,} \\
	\end{array} \right. \\
	& \left\{ \begin{array}{l}
	b(\vec{k},s)\text{ annihilates a positive energy fermion,} \\
	b^{\dagger}(\vec{k},s)\text{ creates a positive energy fermion,} \\
	d(\vec{k},s)\text{ annihilates a negative energy fermion,} \\
	d^{\dagger}(\vec{k},s)\text{ creates a negative energy fermion.}
	\end{array} \right.
\end{align*}

\noindent In the ordinary method, recalling that the Dirac sea is the true fermion vacuum, we can use $d^{\dagger}$ as the creation operator and $d$ as the annihilation operator for negative energy fermions. Then, the operators $d^{\dagger}$ and $d$ are re-interpreted as the annihilation operator and creation operator for positive energy holes. In this manner, we obtain the particle picture in the real world. In this procedure, negative energy fermions are regarded as actually existing entities. 

For bosons, in contrast to the fermions, we rewrite (15) as 

\begin{align}
	& \> a_{i-}\equiv \tilde{a}_{i-}^{\dagger}, \quad a_{i-}^{\dagger}
	\equiv \tilde{a}_{i-}, \\
	& \left[\tilde{a}_{i-}(\vec{k}),\tilde{a}_{j-}^{\dagger}
	(\vec{k}^{\prime})\right]=+\delta_{ij}\delta^3
	(\vec{k}-\vec{k}^{\prime}).
\end{align}

\noindent This implies that we can treat negative energy bosons in the same manner as positive energy bosons. Consequently, the true vacua for positive and negative energy bosons, which are denoted $||0_+^{\text{ordinary}}\rangle$ and $||0_-^{\text{ordinary}}\rangle$, respectively\footnotemark, are given by 

\footnotetext{In the following, we denote the vacua by, for example in the boson case, $|0_{\pm}\rangle$ in the system of single particle, and $||0_{\pm}\rangle$ in the system with many particles.}

\begin{align*}
	& a_{i+}||0_+^{\text{ordinary}}\rangle =0, \\
	& \tilde{a}_{i-}||0_-^{\text{ordinary}}\rangle =0. \tag{$18^{\prime}$}
\end{align*}

\noindent Thus, in the true vacuum, meaning the one on which our experimental world is built, both the negative and positive energy vacua are empty when using the particle $a_{i+}$ and anti-particle $\tilde{a}_{i-}$ annihilation operators respectively. However, in order to have a supersymmetry relation to the analogous negative energy states for the fermions, we would have liked to consider, instead of $||0_-^{\text{ordinary}}\rangle$, a vacuum so that it were empty with respect to the negative energy bosons described by $a_{i-}$ and $a_{i-}^{\dagger}$. That is to say we would have liked a vacuum obeying $a_{i-}||0_-^{\text{wanted}}\rangle =0$. Because of ($18^{\prime}$) it is, however, immediately seen that this $||0_-^{\text{wanted}}\rangle$ cannot exist.

Perhaps the nicest way of describing this extension is by means of the double harmonic oscillator to be presented in Section 3 below, but let us stress that all we need is a formal extrapolation to also include the possibility of negative numbers of bosons.

\subsection{Supersymmetry invariant vacuum}

\vspace{0.5cm}

As described in Subsection 2.1, when considered in terms of supersymmetry, there is a difference between the boson and fermion pictures. In the present subsection, we give preliminary considerations to the problem determining the nature of a boson sea that would correspond to the Dirac sea for the fermion case. To this end, we impose the natural condition within the supersymmetric theory that the vacuum be supersymmetry invariant. 

We first rewrite the supersymmetry charges $Q_i$ derived from the supersymmetry currents described by Eq.(11) in terms of the creation and annihilation operators as 

\begin{align}
	Q_i & =\int d^3\vec{x} J_i^0(x) \nonumber \\
	& =i\int d^3\vec{k} \sum_{s=\pm} 
	\left\{b(\vec{k},s)u(\vec{k},s)
	a_{i+}^{\dagger}(\vec{k})-d(\vec{k},s)v(\vec{k},s)a_{i-}^{\dagger}
	(\vec{k})\right\}, \\
	\bar{Q}_i & =\int d^3\vec{x} \bar{J}_i^0(x) \nonumber \\
	& =-i\int d^3\vec{k} \sum_{s=\pm} 
	\left\{b^{\dagger}(\vec{k},s)
	\bar{u}(\vec{k},s)a_{i+}(\vec{k})-d^{\dagger}(\vec{k},s)\bar{v}
	(\vec{k},s)a_{i-}(\vec{k})\right\}.
\end{align}

\noindent By applying these charges, the condition for the vacuum to be supersymmetric can be written 

\begin{align}
	Q_i||0\rangle=\bar{Q}_i||0\rangle=0.
\end{align}

\noindent We then decompose the total vacuum into the boson and fermion vacua, $||0_{\pm}\rangle$ and $||\tilde{0}_{\pm}\rangle$, writing 

\begin{align}
	||0\rangle \equiv ||0_+\rangle \otimes ||0_-\rangle \otimes 
	||\tilde{0}_+\rangle \otimes ||\tilde{0}_-\rangle ,
\end{align}

\noindent where $\otimes$ denotes the direct product, and $||\tilde{0}_-\rangle$ is the Dirac sea, given by 

\begin{align*}
	||\tilde{0}_-\rangle =\bigg\{\prod_{\vec{p},s}d^{\dagger}
	(\vec{p},s)\bigg\}||\tilde{0}\rangle . \tag{$23^{\prime}$}
\end{align*}

\noindent The fermion vacuum described by the above two equations consists of $||\tilde{0}_+\rangle$ and $||\tilde{0}_-\rangle$. Here, $||\tilde{0}_+\rangle$ represents an empty vacuum, annihilated by the ordinary $b$ operator, while $||\tilde{0}_-\rangle$, given by Eq.($23^{\prime}$), represents the Dirac sea, which is obtained through application of all $d^{\dagger}$. The condition for the bosonic vacuum reads 

\begin{align}
	& a_{i+}(\vec{k})||0_+\rangle =0, \\
	& a_{i-}^{\dagger}(\vec{k})||0_-\rangle =0.
\end{align}

\noindent It is evident that the vacuum of the positive energy boson $||0_+\rangle$ is the empty one, vanishing under the annihilation operator $a_{i+}$. On the other hand, the vacuum of the negative energy boson $||0_-\rangle$ is defined such that it vanishes under the operator $a_{i-}^{\dagger}$ that creates the negative energy quantum. This may seem very strange. 

One could call the strange algebra (15) and (25)  a ``negative energy sector" of the negative energy particle states, contrary to the more usual creation and annihilation systems (14) and (24) which could rather be called ``positive energy sector" of the positive ones.

In the next section, using the fact that the algebras (14) and (15) constitute that is essentially a harmonic oscillator system with infinitely many degrees of freedom, we investigate in detail the vacuum structure by considering the simplest one-dimensional harmonic oscillator system. In fact, we will find the explicit form of the vacuum $||0_-\rangle$ that is given by a coherent state of the excited states of all the negative energy bosons.

\section{Double harmonic oscillator}

\vspace{0.5cm}

When looking for solutions to the Klein-Gordon equation for energy (and momentum) it is well-known that, we must consider not only the positive energy particles but also the negative energy ones. In the previous section, we found that in order to implement the analogy to the Dirac sea for fermions suggested by supersymmetry, we would have liked to have at our disposal the possibility to organize an analogy of the filling of the Dirac sea (for fermions). This should be organized so as to go from a sector with the naive algebra for $a_{i-}$ meaning as $a_{i-}||0_-^{\text{wanted}}\rangle =0$ to an analogy of the filled Dirac sea being a negative energy sector $a_{i-}^{\dagger}||0_-\rangle =0$. However, so far we are familiar in the description of boson state filling by means of a harmonic oscillator with a positive energy sector. In the present section we introduce the concept of a negative energy sector as an extension of the harmonic oscillator spectrum to a negative energy\footnotemark. Thereby we have to extend the ordinary meaning of the wave function (in this case for the harmonic oscillator). Performing this, we find that the vacuum of the negative energy sector leads to a ``boson sea", corresponding to the Dirac sea of fermions.

\footnotetext{See Ref.~\cite{holten} for another negative energy solution of (deformed) harmonic oscillator.}

\subsection{Analytic wave function of the harmonic oscillator}

\vspace{0.5cm}

As is well known, the eigenfunction $\phi (x)$ of a one-dimensional Schr\"{o}dinger equation in the usual treatment should satisfy the square integrability condition, 

\begin{align}
	\int_{-\infty}^{+\infty}dx \> |\phi (x)|^2<+\infty .
\end{align}

\noindent If we apply this condition to a one-dimensional harmonic oscillator, we obtain as the vacuum solution only the empty one satisfying (24), 

\begin{align*}
	a_+|0\rangle =0.
\end{align*}

\noindent Thus, we are forced to extend the condition for physically allowed wave functions in order to obtain ``boson sea" analogous to the Dirac sea. In fact we extend the condition (26), replacing it by the condition under which, when we analytically continue $x$ to the entire complex plane, the wave function $\phi (x)$ is analytic and only an essential singularity is allowed as $|x|\! \to \! \infty$.

The one-dimensional harmonic oscillator is given by 

\begin{align}
	\left( -\frac{1}{2}\frac{d^2}{dx^2}+\frac{1}{2}x^2 \right) 
	\phi (x)=E\phi (x).
\end{align}

\noindent Under the condition that the wave function $\phi (x)$ be analytic, the function $H_n(x)e^{-\frac{1}{2}x^2}$ is obviously a solution, where $H_n(x)$ denotes the Hermite polynomial. In order to find further solutions we assume the form 
\begin{align}
	\phi (x)=f(x)e^{\pm \frac{1}{2}x^2}, 
\end{align}

\noindent where $f(x)$ is determined below. Inserting (28) into (27), we obtain the following equation for $f(x)$: 

\begin{align}
	\frac{f^{\prime \prime}(x)}{f(x)}\pm \frac{2xf^{\prime}(x)}{f(x)}=
	-2\left(E\pm \frac{1}{2}\right).
\end{align}

\noindent We then assume that for large $|x|$ the second term on the left-hand side dominates the first term. With this, we obtain from Eq.(29) the simpler equation 

\begin{align}
	\frac{d\log f(x)}{d\log x}=\mp E-\frac{1}{2}.
\end{align}

\noindent The right-hand side of the Eq.(30) is constant, and we denote it by $n\in \mathcal{R}$: 

\begin{align}
	n\equiv \mp E-\frac{1}{2}\in \mathcal{R}.
\end{align}

\noindent We thus find that the large $|x|$ behavior of $f(x)$ is given by 

\begin{align}
	f(x)\sim x^n.
\end{align}

\noindent If $n$ is negative or a non-integer then $f(x)$ may have a cut when $x$ is analytically continued to the whole complex plane. This contradicts our assumption. Therefore, it should be the case that 

\begin{align*}
	n\in \mathcal{Z}_+\cup \{0\}; 
\end{align*}

\noindent i.e. $n$ is a positive integer or zero ($n=0,1,2,\cdots$). Thus, the energy spectrum is given by 

\begin{align}
	E=\mp n\pm \frac{1}{2}=\mp \left( n+\frac{1}{2}\right) , \qquad 
	n\in \mathcal{Z}_{+}\cup \{0\}.
\end{align}

\noindent In view of the above arguments regarding the energy spectrum, the eigenfunction for a negative energy $E$, i.e. for the upper sign in (33), can be written 

\begin{align*}
	\phi (x)=f(x)e^{+\frac{1}{2}x^2}.
\end{align*}

\noindent We determine the explicit form of $f(x)$ in the following: As is well known, the ordinary harmonic oscillator is characterized by the energy eigenvalues (33) with the lower sign and the eigenfunctions 

\begin{align}
	& E=n+\frac{1}{2},\qquad n=0,1,2,\cdots , \\
	& \phi_n(x)=\left( \sqrt{\pi}2^nn!\right)^{-\frac{1}{2}}H_n(x)
	e^{-\frac{1}{2}x^2}.
\end{align}

\noindent Now, we see that the negative energy sector is described by Eq.(33) with the upper sign: 

\begin{align}
	E=-\left( n+\frac{1}{2}\right) ,\qquad n=0,1,2,\cdots .
\end{align}

\noindent The negative energy sector eigenfunctions, although satisfying the same equation [i.e., Eq.(24)] as the positive energy eigenfunctions, are found to be 

\begin{align*}
	\phi (x)=f(x)e^{+\frac{1}{2}x^2}.
\end{align*}

\noindent This implies that the harmonic oscillator in the negative energy sector is formally described by simply replacing $x$ by $ix$ in all the results for the positive energy sector. Thus, the eigenfunctions in the negative energy sector can be obtained through the replacement 

\begin{align}
	& \left( -\frac{1}{2}\frac{d^2}{d(ix)^2}+\frac{1}{2}(ix)^2 \right) 
	\phi (x)=E\phi (x) \nonumber \\
	& \quad \Longleftrightarrow \left( -\frac{1}{2}\frac{d^2}{dx^2}
	+\frac{1}{2}x^2 \right) \phi (x)=-E\phi (x).
\end{align}

\noindent Therefore the solutions read 

\begin{align}
	& E=-\left( n+\frac{1}{2}\right) ,\qquad n=0,1,2,\cdots \\
	& \phi_{-n}(x)=A_nH_n(ix)e^{-\frac{1}{2}(ix)^2}=A_nH_n(ix)
	e^{+\frac{1}{2}x^2},
\end{align}

\noindent where $A_n$ denotes a normalization factor. It should be noted that the Hermite polynomial $H_n(x)$ is either an even or odd function of $x$, so that $H_n(ix)$ in Eq.(39) is either purely real or purely imaginary. In the case that it is purely imaginary, the overall factor $i$ of $H_n(ix)$ can be absorbed into the normalization factor $A_n$. Thus, the eigenfunctions of the negative energy sector $\phi _{-n}(x)$ can be treated as a real function. 

\subsection{Representation of the double harmonic oscillator}

\vspace{0.5cm}

In this subsection, we describe in detail how all the states are created by action of the creation and annihilation operators to each vacuum of positive and negative energy sectors. We then proceed to defining explicitly each norm of these sectors and construct total Hilbert space of the double harmonic oscillator.

First of all, we notice that the positive energy sector is exactly the same as that of the ordinary harmonic oscillator. On the other hand, the eigenfunctions are simply obtained by a replacement of the argument $x\! \to \! ix$. However this replacement leads to the transformations of the creation and annihilation operators and their commutation relation as 

\begin{align}
	a_+=\frac{1}{\sqrt{2}}\left( x+\frac{d}{dx}\right) & \to 
	\frac{i}{\sqrt{2}}\left( x-\frac{d}{dx}\right)\equiv a_-^{\dagger}, 
	\nonumber \\
	a_+^{\dagger}=\frac{1}{\sqrt{2}}\left( x-\frac{d}{dx}\right) & \to 
	\frac{i}{\sqrt{2}}\left( x+\frac{d}{dx}\right)\equiv a_-, \\
	& \nonumber \\
	[a_+,a_+^{\dagger}]=+1 & \to [a_-,a_-^{\dagger}]=-1.
\end{align}

\noindent Here, $a_+$ and $a_+^{\dagger}$ are ordinary operators in the positive energy sector which annihilate and create positive energy quanta respectively, and $a_-$ and $a_-^{\dagger}$ are operators in the negative energy sector which, as we will see below, create and annihilate negative energy quanta respectively. In the following, we will omit the irrelevant factors $\frac{1}{\sqrt{2}}$ in (40) and the overall factors of the eigenfunctions. At this stage, we can explicitly check the unusual property mentioned above of the negative energy sector, as explained below. The exotic properties that the vacuum is, in our usual words, annihilated by the creation operator $a_-^{\dagger}$ and the annihilation operator $a_-$ excites states are confirmed as 

\begin{align}
	\text{(vacuum)} \quad & a_-^{\dagger}\phi _{-0}(x)
	=i\left( x-\frac{d}{dx}\right)e^{+\frac{1}{2}x^2}=0, \\
	\text{(excitations)} \quad 
	& \phi_{-1}(x)=2ixe^{+\frac{1}{2}x^2}=a_-\phi _{-0}(x)
	=i\left( x+\frac{d}{dx}\right)e^{+\frac{1}{2}x^2}, \nonumber \\
	& \phi_{-2}(x)=-(4x^2+2)e^{+\frac{1}{2}x^2}=
	a_-\phi _{-1}(x)=-\left( x+\frac{d}{dx}\right)^2e^{+\frac{1}{2}x^2}, 
	\nonumber \\
	& \quad \vdots
\end{align}

\noindent Also, in general, we can confirm that 

\begin{align}
	& \phi_{-n-1}(x)=a_-\phi_{-n}(x), \nonumber \\
	& \phi_{-n+1}(x)=a_-^{\dagger}\phi_{-n}(x), \quad \text{for}\quad 
	n\geq 1. 
\end{align}

\noindent Namely, we can consider that $a_-$ creates a negative energy quantum and $a_-^{\dagger}$ annihilates a negative energy quantum. 

Here some notices to the readers are in order. One would think that in order to transform $a_-$ and $a_-^{\dagger}$ into each others under the hermite conjugation as eq. (40), the replacement $x\! \to \! ix$ would be needed to make the hermite operators $\hat{x}=x$ and $\hat{p}=-i\frac{d}{dx}$ transform to anti-hermite operators. However, in our formalism, the operators $\hat{x}$ and $\hat{p}$ must be forbidden to be anti-hermitean in the following reason: If $\hat{x}=x$ were anti-hermitean, $x$ would turn out to be pure imaginary. Then the eigenfunction (39) in the negative energy sector coincides with that of the positive energy sector eq. (35), and thus the vacuum in the negative energy sector coincides with that of the positive one. This implies that the vacuum in the negative energy sector is also empty which is, needless to say, unwanted. Therefore we should treat $\hat{x}$ and $\hat{p}$ in the negative energy sector as hermitean operator and therefore the replacement $x\! \to \! ix$ should be understood as a formal replacement for deriving the eigenfunction in the negative energy sector.

Unfortunately, this is not the end of the story: This interpretation, i.e. $\hat{x}$ and $\hat{p}$ be hermitean induces a new problem. 

The problem is, from the result in Section 2, the creation and annihilation operators $a_-$ and $a_-^{\dagger}$ should satisfy 

\begin{align*}
	& [a_-,a_-^{\dagger}]=-1, \tag{$15^{\prime}$} \\
	& a_-^{\dagger}|0_-\rangle =0. \tag{$25^{\prime}$}
\end{align*}

\noindent However, in order that the vacuum $|0_-\rangle$ in the negative energy sector should be annihilated by $a_-^{\dagger}$, the operators $a_-$ and $a_-^{\dagger}$ should be represented as 

\begin{align*}
	a_-=l\left( x+\frac{d}{dx}\right), \quad 
	a_-^{\dagger}=l^{\ast}\left( x-\frac{d}{dx}\right), 
\end{align*}

\noindent in terms of a certain constant $l$. The point is that $a_-$ and $a_-^{\dagger}$ are ordinarily hermite conjugate each other, but since 

\begin{align*}
	[a_-,a_-^{\dagger}]=-1 \quad \Longleftrightarrow \quad ll^{\ast}=-1, 
\end{align*}

\noindent we are forced to conclude that there cannot exist such a representation of $a_-$ and $a_-^{\dagger}$. 

From the above consideration, the expression (40) is too naive and not correct one: We definitely need to extend our formulation in one way or another. In the following we propose our extension in detail.

At the starting point we interpret the replacement $x\! \to \! ix$, as a matter of fact, 

\begin{align*}
	1\otimes x\! \to \! i\otimes x,
\end{align*}

\noindent and then rewrite eq. (40) as 

\begin{align}
	a_+=1\otimes \frac{1}{\sqrt{2}}\left( x+\frac{d}{dx}\right) & \to 
	i\otimes \frac{1}{\sqrt{2}}\left( x-\frac{d}{dx}\right)
	\equiv a_-^{\dagger}, 
	\nonumber \\
	a_+^{\dagger}=1\otimes \frac{1}{\sqrt{2}}\left( x-\frac{d}{dx}\right)
	 & \to i\otimes \frac{1}{\sqrt{2}}\left( x+\frac{d}{dx}\right)
	\equiv a_-.
\end{align}

\noindent This implies that the coordinate $x$ should in reality be written as $r\otimes x$, where $r\! =\! \pm 1,\pm i$ is a degree of freedom to label the positive or negative energy sector. In this notation, the hermitean conjugation of the creation and annihilation operators is defined by 

\begin{align}
	\{a_-\}^{\dagger}=\left\{i\otimes \frac{1}{\sqrt{2}}
	\left( x+\frac{d}{dx}\right)\right\}^{\dagger}\equiv i\otimes 
	\left\{\frac{1}{\sqrt{2}}\left( x+\frac{d}{dx}\right)\right\}^{\dagger}
	=a_-^{\dagger}.
\end{align}

\noindent This means that the hermitean conjugation in ordinary sense does not act to the label $r$ for distinguishing the energy sector. That is, due to the exclusion of the imaginary unit $i$ from the hermitean conjugation of $a_-$ and $a_-^{\dagger}$, we obtain $[a_-,a_-^{\dagger}]=-1 \Leftrightarrow r^2=-1$, and then we can represent the algebra (15$^{\prime}$). As far as the hermiticity is kept, the following map in the algebra holds 

\begin{align}
	r\otimes \hat{O}=1\otimes r\hat{O}\simeq r\hat{O}, \quad 
	\text{only for } \> r=\pm 1.
\end{align}

\noindent The case where the hermiticity is broken is that for $r=\pm i$ and thus e.g. for Eq. (46). 

On the other hand, in the Hilbert space, the replacement $1\otimes x\! \to \! i\otimes x$ in practice means that e.g. the eigenfunction of the vacuum $\phi_{+0}(x)$ is transformed as 

\begin{align}
	& \phi_{+0}(x)=e^{-\frac{1}{2}x^2}\simeq 1\otimes e^{-\frac{1}{2}x^2}
	=e^{-\frac{1}{2}(1\otimes x)^2} \nonumber \\
	& \quad \longrightarrow e^{-\frac{1}{2}(i\otimes x)^2}
	=e^{1\otimes (+\frac{1}{2}x^2)}\simeq e^{+\frac{1}{2}x^2}=\phi_{-0}(x).
\end{align}

\noindent The map corresponding to the algebra (47) is given by 

\begin{align}
	r\otimes \zeta (x)=1\otimes r\zeta (x)\simeq r\zeta (x), \quad 
	\text{for } \> r=\pm 1,\pm i,
\end{align}

\noindent which is nothing but the ordinary function. This property is confirmed by the fact that all the eigenfunctions are even or odd function of $x$. Furthermore the hermitean conjugation in the Hilbert space is defined as the ordinary complex conjugation 

\begin{align}
	\left\{r\otimes \phi_n\right\}^{\dagger}(x)
	=r^{\ast}\otimes \left\{\phi_n(x)\right\}^{\ast}
	\simeq r^{\ast}\left\{\phi (x)\right\}^{\ast}, \quad 
	n=\pm 0,\pm 1,\pm 2,\cdots.
\end{align}

\noindent We defined so in connection with the definition of the norm yet to be defined in the next subsection. Thus the replacement $1\otimes x\! \to \! i\otimes x$ is consistent for all the excited states by defining it as the same manner as eq. (48).

From the above facts, in the remainder of the paper, we can rewrite 

\begin{align*}
	& a\equiv a_+=-ia_-=(x+\frac{d}{dx}), \\
	& a^{\dagger}\equiv a_+^{\dagger}=-ia_-^{\dagger}=(x-\frac{d}{dx}), 
\end{align*}

\noindent since the imaginary unit $i$ is not essential in our argument.

It would be useful to summarize the various results obtained to this point in operator form. We write each vacuum and excited state in the positive and negative energy sectors, respectively, as 

\begin{align}
	& \phi_{+0}(x)=e^{-\frac{1}{2}x^2}\simeq |0_+\rangle , \\
	& \phi_{-0}(x)=e^{+\frac{1}{2}x^2}\simeq |0_-\rangle , \\
	& \phi_n(x)\simeq |n\rangle , \qquad n\in \mathcal{Z}-\{0\}.
\end{align}

\noindent The actions of each operator $a$ and $a^{\dagger}$ acting on various states are summarized in Table 1.

\begin{center}
	\begin{tabular}{|l|l|l|} \hline
	& \multicolumn{2}{|c|}{$n=0,1,2,\cdots$} \\ \hline
	& positive energy sector & negative energy sector \\ \hline
	spectrum & $E=+(n+\frac{1}{2})$ & $E=-(n+\frac{1}{2})$ \\
	& $a^{\dagger}|n\rangle =|n+1\rangle$ & $a|-n\rangle 
	=|-n-1\rangle$ \\
	& $a|n+1\rangle =|n\rangle$ & $a^{\dagger}|-n-1\rangle 
	=|-n\rangle$ \\
	vacuum & $a|0_+\rangle =0$ & $a^{\dagger}|0_-\rangle =0$ \\ \hline
	\multicolumn{3}{c}{} \\
	\multicolumn{3}{c}{Table 1: The actions of $a=(x+\frac{d}{dx})$ 
	and $a^{\dagger}=(x-\frac{d}{dx})$} \\
	\multicolumn{3}{c}{on various states.$\qquad \qquad \qquad \qquad 
	\quad$} \\
	\end{tabular}
\end{center}

\begin{figure}[h]
	\begin{center}
	\begin{picture}(360,25)
	\put(40,10){$\cdots$}
	\put(65,13){$\xrightarrow[]{a^{\dagger}}$}
	\put(65,4){$\xleftarrow[\> a \>]{}$}
	\put(85,8){$|\! -\! 1\rangle$}
	\put(115,13){$\xrightarrow[]{a^{\dagger}}$}
	\put(115,4){$\xleftarrow[\> a \>]{}$}
	\put(135,8){$|0_-\rangle$}
	\put(160,13){$\xrightarrow[]{a^{\dagger}}$}
	\put(180,8){$0$}
	\put(190,4){$\xleftarrow[\> a \>]{}$}
	\put(210,8){$|0_+\rangle$}
	\put(235,13){$\xrightarrow[]{a^{\dagger}}$}
	\put(235,4){$\xleftarrow[\> a \>]{}$}
	\put(255,8){$|\! +\! 1\rangle$}
	\put(285,13){$\xrightarrow[]{a^{\dagger}}$}
	\put(285,4){$\xleftarrow[\> a \>]{}$}
	\put(310,10){$\cdots$.}
	\end{picture}
	\end{center}
	\caption[sequence of states]{Sequence of states}
	\label{sequ}
\end{figure}
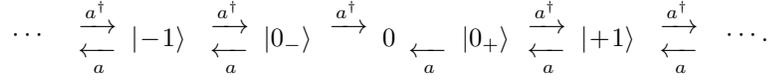%

The important point here is that there exists a gap between the positive and negative energy sectors. Suppose that we write the states in order of their energies as depicted in Fig.~\ref{sequ}.

\noindent As usual, the operators causing transitions in the right and left directions are $a^{\dagger}$ and $a$, respectively. However, between the two vacua $|0_-\rangle$ and $|0_+\rangle$ there is a ``wall" of the classical number $0$, and due to its presence, these two vacua cannot be transformed into each other under the operations of $a$ and $a^{\dagger}$, because the states $|0_{\pm}\rangle$ cannot be obtained by means of acting the operators $a^{\dagger}$ and $a$ on the classical number $0$. In going to the second quantized theory with interactions, there appears to be the possibility of such a transition. However, as was discussed in detail in Subsection 3.5 and Section 4, it turns out that the usual polynomial interactions do not induce such a transition.

\subsection{Inner product of states}

\vspace{0.5cm}

To construct the Hilbert space of the double harmonic oscillator, let us define the inner product of states. As explained above, there exists a gap such that no transition between the two sectors can take place. Thus, we can define the inner product of only states in the same sector. The inner product that in the positive energy sector provides the normalization condition is, as usual, given by 

\begin{align}
	\langle n|m\rangle & \equiv \int_{-\infty}^{+\infty}dx\> 
	\phi_n^{\dagger}(x)\phi_m(x)=\delta_{nm}, \qquad n,m=0_+,1,2,\cdots .
\end{align}

\noindent However, the eigenfunctions in the negative energy sector are obtained as Eq.(39), and consequently a definition of the inner product similar to Eq.(54) leads to divergence in this case if we choose the integration region as $-\infty <x<+\infty$. Therefore we need to define an inner product of a finite value in the negative energy sector.

The simplest candidate of such an inner product may be the following: Since the eigenfunction of the negative energy sector, $H_n(ix)e^{+\frac{1}{2}x^2}$, should be finite, we would like to analytically continue the eigenfunction into the whole complex $x$ plane and integrate along the imaginary axis, 

\begin{align}
	\langle n|m\rangle \equiv \int_{-i\infty}^{+i\infty}dx\> 
	\phi_n^{\dagger}(x)\phi_m(x)<+\infty, \qquad n,m=0_-,-1,-2,\cdots.
\end{align}

\noindent Eq. (55) implies that the function $e^{+\frac{1}{2}x^2}$ which is divergent on the real axis is analytically continued so as to be convergent onto the imaginary axis. Essentially in this manner we treat $e^{+\frac{1}{2}x^2}$ as $e^{-\frac{1}{2}x^2}$, which indicates that the wave function in the negative energy sector employs as that of the positive one. Therefore we may face the same problem as the replacement $x\! \to \! ix$ which was discussed in the previous subsection and thus the method described so far is failed. However there is a remedy of our ill defined method to reactivate as a definition of the inner product. If we consider only the negative energy sector alone, the eigenfunction is convergent only on the imaginary axis of the entire complex $x$ plane, more rigorously within two regions $-\frac{\pi}{4}<\arg x<\frac{\pi}{4}$ and $\frac{3\pi}{4}<\arg x<\frac{5\pi}{4}$. In any way, we face the same problem described above if we choose these convergent regions and thus fail to define the right norm.

From these considerations we would like to take a viewpoint that it is sufficient to define the norm of the negative energy sector as the 2 dimensional harmonic oscillator, since the problem of the negative energy solution occurs only when we consider the complex scalar field theories.

Then, the Schr\"{o}dinger equation of the 2 dimensional harmonic oscillator reads 

\begin{align}
	\left\{\left( -\frac{1}{2}\frac{d^2}{dx^2}+\frac{1}{2}x^2 \right)
	+\left( -\frac{1}{2}\frac{d^2}{dy^2}+\frac{1}{2}y^2 \right)\right\}
	\phi (x,y)=E\phi (x,y).
\end{align}

\noindent There are three kinds of the solutions. Hereafter we abbreviate the overall factors of the wave functions.

\begin{align}
	\text{i) (pos.)} \otimes \text{(pos.)} \qquad 
	& \phi_{+n,+m} (x,y)=H_n(x)e^{-\frac{1}{2}x^2}
	H_m(y)e^{-\frac{1}{2}y^2}, \nonumber \\
	& E_{+n,+m}=\left(n+\frac{1}{2}\right)+\left(m+\frac{1}{2}\right)
	=n+m+1, \\
	\text{ii) (pos.)} \otimes \text{(neg.)} \qquad 
	& \phi_{+n,-m} (x,y)=H_n(x)e^{-\frac{1}{2}x^2}
	H_m(iy)e^{+\frac{1}{2}y^2}, \nonumber \\
	& E_{+n,-m}=\left(n+\frac{1}{2}\right)-\left(m+\frac{1}{2}\right)
	=n-m, \\
	\text{iii) (neg.)} \otimes \text{(neg.)} \qquad 
	& \phi_{-n,-m} (x,y)=H_n(ix)e^{+\frac{1}{2}x^2}
	H_m(iy)e^{+\frac{1}{2}y^2}, \nonumber \\
	& E_{-n,-m}=-\left(n+\frac{1}{2}\right)-\left(m+\frac{1}{2}\right)
	=-n-m-1,
\end{align}

\noindent Here $n,m=0,1,2,\cdots$, and (pos.) and (neg.) denote the positive and negative energy sectors respectively. Hereby a comment is in order. One might think that there would be a fourth kind of solution of the form\footnotemark 

\begin{align*}
	\phi (x,y)=f(x,y)e^{\pm xy}.
\end{align*}

\noindent This may look strange because it does not contain the ``Gaussian" factor like $e^{\pm \frac{1}{2}x^2}$. However, if we make the change of coordinates as $x=\frac{1}{\sqrt{2}}(u+v)$ and $y=\frac{1}{\sqrt{2}}(u-v)$, we find out that this solution is exactly equivalent to the second one ii), because the Schr\"{o}dinger equation (56) is transformed to 

\begin{align}
	\left\{\left( -\frac{1}{2}\frac{d^2}{du^2}+\frac{1}{2}u^2 \right)
	+\left( -\frac{1}{2}\frac{d^2}{dv^2}+\frac{1}{2}v^2 \right)\right\}
	\phi (u,v)=E\phi (u,v), \nonumber
\end{align}

\noindent and, for example, the exponential term $e^{\pm xy}$ becomes $e^{\pm \frac{1}{2}(u^2-v^2)}$. Therefore, this fourth case of the solutions is not independent so that we consider only the three solutions i)-iii).

\footnotetext{We would like to thank Roman Jackiw for his presentation of this solution in private communication. For details of this solution, see Refs.~\cite{jackiw,thooft}.}

In order to find the convergent regions of the inner product, we introduce a complex coordinate $z=x+iy$ where, needless to say, $x$ and $y$ are real. In terms of $z$ the three solutions are rewritten as 

\begin{align*}
	\text{i) (pos.)} \otimes \text{(pos.)} \qquad 
	& \phi_{+n,+m} (x,y)=h_{+n,+m}(z,\bar{z})e^{-\frac{1}{2}z\bar{z}}, \\
	& E_{+n,+m}=\left(n+\frac{1}{2}\right)+\left(m+\frac{1}{2}\right)
	=n+m+1, \tag{$57^{\prime}$} \\
	\text{ii) (pos.)} \otimes \text{(neg.)} \qquad 
	& \phi_{+n,-m} (x,y)=h_{+n,-m}(z,\bar{z})
	e^{-\frac{1}{4}(z^2+\bar{z}^2)}, \\
	& E_{+n,-m}=\left(n+\frac{1}{2}\right)-\left(m+\frac{1}{2}\right)
	=n-m, \tag{$58^{\prime}$} \\
	\text{iii) (neg.)} \otimes \text{(neg.)} \qquad 
	& \phi_{-n,-m} (x,y)=h_{-n,-m}(z,\bar{z})e^{+\frac{1}{2}z\bar{z}}, \\
	& E_{-n,-m}=-\left(n+\frac{1}{2}\right)-\left(m+\frac{1}{2}\right)
	=-n-m-1. \tag{$59^{\prime}$}
\end{align*}

\noindent In these expressions the functions $h_{\pm n,\pm m}(z,\bar{z})$ are those of the product of the two Hermite polynomials in $(57)\sim(59)$ and are the polynomials in terms of $z$ and $\hat{z}$. The important point is in their exponential factors because they decide the convergence properties.

In the first case i) ``$\text{(pos.)} \otimes \text{(pos.)}$" solution $(57)(57^{\prime})$ is just the solution of the ordinary 2 dimensional harmonic oscillator and represents the complex scalar field decomposed into two real scalar fields as $A(x)=F(x)+iG(x)$ instead of (12). This solution describes the two empty boson vacua and is not supersymmetric to the solution of Dirac field including the Dirac sea. The norm is ordinarily defined as 

\begin{align}
	\text{i) (pos.)} & \otimes \text{(pos.)} \nonumber \\
	& \text{for } n,n^{\prime},m,m^{\prime}=+0,1,2,\cdots, \nonumber \\
	& \langle n,m|n^{\prime},m^{\prime}\rangle 
	=\int_{-\infty}^{+\infty}dx \int_{-\infty}^{+\infty}dy \> 
	\phi_{n,m}^{\dagger} (x,y)\phi_{n^{\prime},m^{\prime}} (x,y)
	=\delta_{n,n^{\prime}}\delta_{m,m^{\prime}}.
\end{align}

\noindent Let us go on the case iii) ``$\text{(neg.)} \otimes \text{(neg.)}$" solution $(59)(59^{\prime})$. The function $e^{+\frac{1}{2}z\bar{z}}$ diverges as $|z|\! \to \! \infty$, irrespective to the arg $z$ and the integration cannot be defined and so the norm does not exist. Thus strictly speaking the case iii) is not the solution.

As the last case we investigate ii) ``$\text{(pos.)} \otimes \text{(neg.)}$" solution $(58)(58^{\prime})$ which is nothing but the solution of the complex scalar field theory we want. The exponential factor $e^{-\frac{1}{4}(z^2+\bar{z}^2)}$ are both Gaussian in $z$ and $\bar{z}$ so that it is convergent in the shaded region $\gamma$ in Fig. \ref{gamma} and thus the integration of the norm is defined in the region $\gamma$ in the following manner: 

\begin{figure}[htb]
	\begin{center}
	\begin{picture}(120,140)
	\put(0,60){\vector(1,0){120}}
	\put(60,0){\vector(0,1){120}}
	\put(120,120){\line(1,0){10}}
	\put(120,120){\line(0,1){10}}
	\put(122,122){$z$}
	\put(122,57.5){Re$\> z=x$}
	\put(50,123){Im$\> z=y$}
	\put(54,46){$0$}
	\put(0,75){$\gamma$}
	\put(5,5){\line(1,1){110}}
	\put(5,115){\line(1,-1){110}}
	\put(70,50){\line(0,1){20}}
	\put(80,40){\line(0,1){40}}
	\put(90,30){\line(0,1){60}}
	\put(100,20){\line(0,1){80}}
	\put(110,10){\line(0,1){100}}
	\put(50,50){\line(0,1){20}}
	\put(40,40){\line(0,1){40}}
	\put(30,30){\line(0,1){60}}
	\put(20,20){\line(0,1){80}}
	\put(10,10){\line(0,1){100}}
	\end{picture}
	\end{center}
	\caption[region]{Region $\gamma$ for which the inner product (61) 
	converges.}
	\label{gamma}
\end{figure}
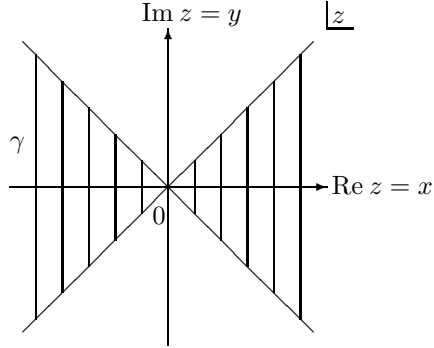%

\begin{align}
	\text{ii) (pos.)} & \otimes \text{(neg.)} \nonumber \\
	& \text{for } n,n^{\prime}=+0,1,2,\cdots, \nonumber \\
	& \quad \> \> m,m^{\prime}=-0,-1,-2,\cdots, \nonumber \\
	& \langle n,m|n^{\prime},m^{\prime}\rangle 
	=\int_{\gamma}dx\> dy \> 
	\phi_{n,m}^{\dagger} (x,y)\phi_{n^{\prime},m^{\prime}} (x,y)
	=\delta_{n,n^{\prime}}\delta_{m,m^{\prime}}, \\
	& \gamma =\left\{ (x,y)\Big|\tan \theta =\frac{y}{x},
	-\frac{\pi}{4}<\theta <\frac{\pi}{4},
	\frac{3\pi}{4}<\theta <\frac{5\pi}{4}\right\}.
\end{align}

\noindent Some comments are in order. As was argued in the previous subsection, the hermite conjugation in the Hilbert space is just defined as ordinary complex conjugation. Thus $\phi_{n,m}^{\dagger} (x,y)\phi_{n^{\prime},m^{\prime}} (x,y)$ is always positive and in terms of polar coordinate 

\begin{align*}
	\int_{\gamma}dx \> dy=\int_{0}^{\infty}dr \left(\int_{-\frac{\pi}{4}}^{\frac{\pi}{4}}d\theta +\int_{\frac{3\pi}{4}}^{\frac{5\pi}{4}}d\theta \right),
\end{align*}

\noindent the integration is always positive like (61) and so normalizable. We thus conclude that the norm of the Hilbert space for the case of (pos.)$\otimes$(neg.) is defined by (61) and (62).

To close this subsection we summarize the conclusions obtained here. Since the norm of the states in the positive and negative energy sectors and thus the Hilbert space is appropriately constructed, excited states in the negative energy sector is expected to describe some physical states. In fact in the later section the boson sea is justified as the filled state of the negative energy particles, just like that of the Dirac sea for fermions.

\subsection{Fermionic harmonic oscillator}

\vspace{0.5cm}

We have derived the explicit form of the eigenfunction in the negative energy sector. However, it should be clarified that the boson vacuum $|0_-\rangle \simeq e^{+\frac{1}{2}x^2}$ in the negative energy sector is really a ``boson sea", corresponding to the Dirac sea in the case of fermions. Here, we recall the concept of a ``sea" by considering the fermionic harmonic oscillator as the first quantized theory.

The one-dimensional fermionic harmonic oscillator is defined in terms of the Grassmann odd operators that satisfy the anti-commutator 

\begin{align}
	\{b,b^{\dagger}\}=1.
\end{align}

\noindent The representation for these operators is expressed in terms of the real Grassmann variable $\theta$ as 

\begin{align}
	b=\theta , \quad b^{\dagger}=\frac{d}{d\theta}.
\end{align}

\noindent This follows from the fact that $b$ and $b^{\dagger}$ are Grassmann odd and they act from the left in the functional space consisting of the Grassmannian functions. The Hamiltonian for the fermionic harmonic oscillator reads 

\begin{align}
	H=\frac{1}{2}[b^{\dagger},b]=b^{\dagger}b-\frac{1}{2}=N-\frac{1}{2}, 
\end{align}

\noindent where $N=b^{\dagger}b$ is the particle number operator in the second quantized theory. The solutions of the Schr\"{o}dinger equation 

\begin{align}
	\left(b^{\dagger}b-\frac{1}{2}\right)|\tilde{n}_+\rangle 
	=E_n|\tilde{n}_+\rangle 
	\Longrightarrow \left(\frac{d}{d\theta}\theta-\frac{1}{2}\right)
	\rho_n(\theta)=E_n\rho_n(\theta)
\end{align}

\noindent have the following forms: 

\begin{align}
	& E_n=n-\frac{1}{2},\qquad n=0,1 \\
	& \rho_0(\theta)=\theta \simeq |\tilde{0}_+\rangle , \\
	& \rho_1(\theta)=1 \simeq |\tilde{1}_+\rangle , \\
	& \text{vacuum} \quad b|\tilde{0}_+\rangle =0.
\end{align}

\noindent Here, the normalization condition 

\begin{align}
	\int d\theta \> \rho_n^{\ast}(\theta)\left(b+b^{\dagger}\right)
	\rho_m(\theta)=\delta_{nm}
\end{align}

\noindent has been used. Note that the vacuum $|0_+\rangle$ is annihilated by the annihilation operator $b$.

In the argument presented in Section 2 for the Dirac sea, in the negative energy sector, the excited states have the same energies as in the positive energy sector when we exchange the operators $b$ and $b^{\dagger}$. With this replacement, the Hamiltonian reads 

\begin{align}
	H=\frac{1}{2}[b,b^{\dagger}]=bb^{\dagger}-\frac{1}{2},
\end{align}

\noindent and consequently the Schr\"{o}dinger equation is changed to 

\begin{align}
	\left(bb^{\dagger}-\frac{1}{2}\right)|\tilde{n}_-\rangle 
	=E_n|\tilde{n}_-\rangle 
	\Longrightarrow \left(\theta \frac{d}{d\theta}-\frac{1}{2}\right)
	\chi_n(\theta)=E_n\chi_n(\theta).
\end{align}

\noindent The solutions of Eq.(73) are given by 

\begin{align}
	& E_n=n-\frac{1}{2},\qquad n=0,1 \\
	& \chi_0(\theta)=1 \simeq |\tilde{0}_-\rangle , \\
	& \chi_1(\theta)=\theta \simeq |\tilde{1}_-\rangle , \\
	& \text{vacuum} \quad b^{\dagger}|\tilde{0}_-\rangle =0,
\end{align}

\noindent where we have used the same normalization condition as for the positive energy sector (71). Here, it should be noted that the vacuum $|\tilde{0}_-\rangle$ in the negative energy sector is annihilated by the creation operator $b^{\dagger}$. This implies that the vacuum $|\tilde{0}_-\rangle$ has precisely the properties postulated to exist for the Dirac sea. In order to understand more precisely the reason why the vacuum $|\tilde{0}_-\rangle$ in the negative energy sector is naturally characterized by the concept of a ``sea", we write the states in terms of the Grassmann variable: 

\begin{align}
	& |\tilde{0}_+\rangle \simeq \theta ,\quad |\tilde{0}_-\rangle 
	\simeq 1 \nonumber \\
	& \Longrightarrow |\tilde{0}_-\rangle =b^{\dagger}|\tilde{0}_+\rangle .
\end{align}

\noindent This equation shows that the vacuum $|\tilde{0}_-\rangle$ in the negative energy sector is obtained by exciting one quantum of the empty vacuum $|\tilde{0}_+\rangle$. If we employ this fact in the second quantized theory, i.e. an infinitely many body system, the vacuum in our picture is equivalent to the Dirac sea.

\subsection{The meaning of boson vacuum in the negative energy sector}

\vspace{0.5cm}

We are ready to clarify the meaning in which the vacuum in the negative energy sector of bosons forms a sea, applying the foregoing arguments to the case of bosons. 

The vacua $|0_+\rangle$ and $|0_-\rangle$ in the positive and negative energy sectors are 

\begin{align}
	& |0_+\rangle \simeq e^{-\frac{1}{2}x^2}, \\
	& |0_-\rangle \simeq e^{+\frac{1}{2}x^2}.
\end{align}

\noindent In order to demonstrate how $|0_-\rangle$ represents a sea, we derive a relation between the two vacua (79) and (80) analogous to that in the fermion case. In fact, by comparing the explicit functional forms of each vacuum, we easily find the relation 

\begin{align}
	& e^{+\frac{1}{2}x^2}=e^{x^2}\cdot e^{-\frac{1}{2}x^2}, \qquad 
	e^{x^2}=e^{\frac{1}{2}(a+a^{\dagger})^2} \nonumber \\
	& \Longrightarrow 
	|0_-\rangle =e^{\frac{1}{2}(a+a^{\dagger})^2}|0_+\rangle 
	=e^{-\frac{1}{2}(a_-+a_-^{\dagger})^2}|0_+\rangle .
\end{align}

\noindent This relation is preferable for bosons for the following reason. In the fermion case, described by (78), due to the exclusion principle, the Dirac sea is obtained by exciting only one quantum of the empty vacuum. Contrastingly, because in the boson case there is no exclusion principle, the vacuum $|0_-\rangle$ in the negative energy sector is constructed as a sea by exciting all even number of quanta, i.e. an infinite number of quanta. 

Although the basic idea of the boson sea is described above, in this form, it is limited to the first quantized theory. We extend our argument to the second quantized theory in the next section. We present a new particle picture employing our result that the true boson vacuum is similar to the true fermion vacuum, where a hole appearing through the annihilation of one negative energy particle is interpreted as an anti-particle.

\section{Boson sea}

\vspace{0.5cm}

In the previous section, we investigated 1 and 2 dimensional harmonic oscillators in detail. These harmonic oscillators describe, in principle, one particle energy state, and it can have a negative number of excitations. We can apply this result to the second quantization for a complex scalar field, which is simply a system of infinite-dimensional harmonic oscillators. The negative energy quantum describes one negative energy particle, and the vacuum in the negative energy sector of the negative energy particle states is described as the state in which an infinite number of negative energy particles fill all states. 

In the present section, we investigate the boson vacuum structure in detail. We find that by appropriately defining the energy-momentum operator through the Noether's theorem, our method yields the result that all physical phenomena are characterized by positive energy, even in the negative energy sector of the negative energy particle states. Thus it provides a picture that is consistent with the ordinary method derived from a re-definition of the operators (18) and (19). It is important that we can extend all the concepts of the 1 and 2 dimensional harmonic oscillators to the infinite-dimensional case, since, as is well known in the second quantization, the number of independent (i.e. non-interacting) harmonic oscillators becomes infinite. 

\vspace{0.5cm}

To begin with, we summarize the results obtained in Section 2 and 3. To simplify the points, we omit the well-known fermion vacuum, the Dirac sea. Furthermore we consider only one boson, $A(x)$, because two complex scalar fields are needed to preserve the supersymmetry, which does not play an essential role in the present section. If we wish to apply the content of the present section to a supersymmetric hypermultiplet, we simply attach the suffices $i=1,2$ to all the scalar fields.

The mode expansion and the commutation relations are given by 

\begin{align}
	& A(x)=\int \frac{d^3\vec{k}}{\sqrt{(2\pi )^32k_0}}\left\{
	a_+(\vec{k})e^{-ikx}+a_-(\vec{k})e^{ikx}\right\}, \\
	& \left[a_+(\vec{k}),a_+^{\dagger}(\vec{k}^{\prime})\right]=
	+\delta^3(\vec{k}-\vec{k}^{\prime}), \nonumber \\
	& \left[a_-(\vec{k}),a_-^{\dagger}(\vec{k}^{\prime})\right]=
	-\delta^3(\vec{k}-\vec{k}^{\prime}).
\end{align}

\noindent As is noted, the commutation relations of the operators for the positive energy particle states, $a_+(\vec{k})$, and the negative energy particle states, $a_-(\vec{k})$, have opposite signs on the right-hand side. The interpretation of these operators and their conjugates are as follows: 

\begin{align*}
	& \left\{ \begin{array}{l}
	a_+(\vec{k})\text{ annihilates a positive energy boson,} \\
	a_+^{\dagger}(\vec{k})\text{ creates a positive energy boson,} \\
	\end{array} \right. \\
	& \left\{ \begin{array}{l}
	a_-(\vec{k})\text{ annihilates a negative energy boson,} \\
	a_-^{\dagger}(\vec{k})\text{ creates a negative energy boson.} \\
	\end{array} \right.
\end{align*}

\noindent The vacuum is expressed as a direct product of those for the positive and negative energy sectors of the positive and negative energy particles, respectively: 

\begin{align}
	||0\rangle_{\text{boson}}=||0_+\rangle \otimes ||0_-\rangle .
\end{align}

\noindent The conditions for $||0_+\rangle$ and $||0_-\rangle$ are given as 

\begin{align}
	& a_+(\vec{k})||0_+\rangle =0, \\
	& a_-^{\dagger}(\vec{k})||0_-\rangle =0.
\end{align}

\noindent It should be noted that the positive energy sector vacuum $||0_+\rangle$ for the positive energy particles vanishes under the action of the annihilation operator $a_+$, and $||0_+\rangle$ turns out to be empty, while the negative energy sector vacuum $||0_-\rangle$ for the negative ones vanishes under the action of the creation operator $a_-^{\dagger}$. This ends the summary.

\vspace{0.5cm}

Firstly, we clarify the properties of the unfamiliar vacuum $||0_-\rangle$ in the negative energy sector for the negative energy particles, using the result of Section 3. To this end, we study the details of the infinite-dimensional harmonic oscillator, which is identical to a system of a second quantized complex scalar field. The representation of the algebra (83) that is formed by $a_+,a_-$ and their conjugate operators is expressed as 

\begin{align}
	& a_+(\vec{k})=1\otimes \left( A_+(\vec{k})
	+\frac{\delta}{\delta A_+(\vec{k})}\right), 
	\quad a_+^{\dagger}(\vec{k})=1\otimes \left( A_+(\vec{k})
	-\frac{\delta}{\delta A_+(\vec{k})}\right), \\
	& a_-(\vec{k})=i\otimes \left( A_-(\vec{k})
	+\frac{\delta}{\delta A_-(\vec{k})}\right),
	\quad a_-^{\dagger}(\vec{k})=i\otimes \left( A_-(\vec{k})
	-\frac{\delta}{\delta A_-(\vec{k})}\right).
\end{align}

\noindent The Hamiltonian and Schr\"{o}dinger equation of this system as the two infinite-dimensional harmonic oscillators expressed by the coordinates $A_{\pm}(\vec{k})$ read 

\begin{align}
	& H=\int \frac{d^3\vec{k}}{(2\pi )^3}\left\{ -\frac{1}{2}
	\frac{\delta^2}{\delta A_+^2(\vec{k})}
	+\frac{1}{2}A_+^2(\vec{k})+\frac{1}{2}
	\frac{\delta^2}{\delta A_-^2(\vec{k})}
	-\frac{1}{2}A_-^2(\vec{k})\right\}, \\
	& H\Phi [A_+,A_-]=E\Phi [A_+,A_-].
\end{align}

\noindent Here, $\Phi [A_+,A_-]$ denotes a wave functional of the wave functions $A_{\pm}(\vec{k})$. We are now able to write an explicit wave functional for the vacua of the positive and negative energy sectors for the positive and negative energy particles respectively: 

\begin{align}
	& ||0_+\rangle \simeq \Phi_{0_+}[A_+]=e^{-\! \frac{1}{2} \int \! 
	\frac{d^3\vec{k}}{(2\pi )^3}A_+^2(\vec{k})}, \\
	& ||0_-\rangle \simeq \Phi_{0_-}[A_-]=e^{+\! \frac{1}{2} \int \! 
	\frac{d^3\vec{k}}{(2\pi )^3}A_-^2(\vec{k})}.
\end{align}

\noindent We can find a relation between the two vacua for the negative energy particle states via Eq.(81): 

\begin{align}
	||0_-\rangle & =e^{\int \! \frac{d^3\vec{k}}{(2\pi )^3}A_-^2(\vec{k})}
	||0_+\rangle \nonumber \\
	& =e^{-\frac{1}{2}\! \int \! \frac{d^3\vec{k}}{(2\pi )^3}
	\left\{ a_-(\vec{k})+a_-^{\dagger}(\vec{k})\right\}^2}||0_+\rangle .
\end{align}

\noindent Here $||0_+\rangle$ denotes the empty vacuum $e^{-\frac{1}{2}\int \! \frac{d^3\vec{k}}{(2\pi )^3}A_-^2(\vec{k})}$ for the negative energy particle states. From this equation, we see that the vacuum $||0_-\rangle$ is a coherent state constructed from the empty vacuum $||0_+\rangle$ of the positive energy sector by creating all the even number negative energy bosons through the action of $a_-^{\dagger}(\vec{k})$. In this sense, $||0_-\rangle$ is the sea in which all the negative energy boson states are filled.

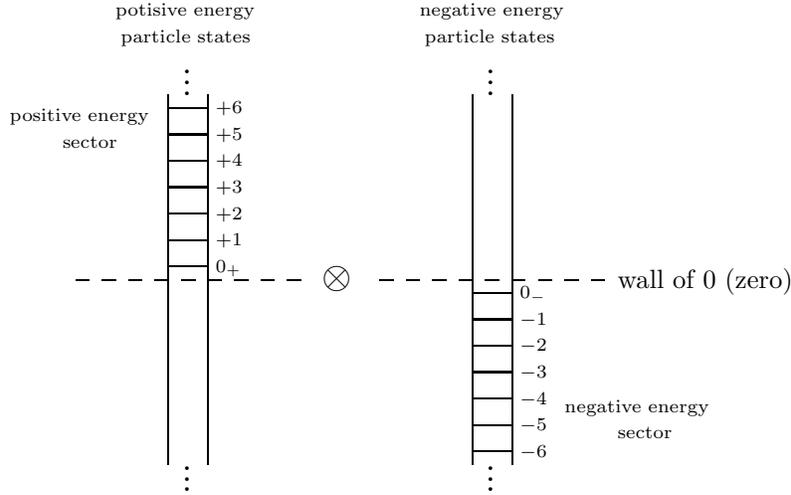
\begin{figure}[ht]
	\begin{center}
	\begin{picture}(200,180)
	\put(35,0){\line(0,1){140}}
	\put(50,0){\line(0,1){140}}
	\put(150,0){\line(0,1){140}}
	\put(165,0){\line(0,1){140}}
	\put(93,67){\Large $\otimes$}
	\put(40,-10){\Large $\vdots$}
	\put(40,140){\Large $\vdots$}
	\put(155,-10){\Large $\vdots$}
	\put(155,140){\Large $\vdots$}
	\put(15,170){\scriptsize potisive energy}
	\put(17,160){\scriptsize particle states}
	\put(130,170){\scriptsize negative energy}
	\put(132,160){\scriptsize particle states}
	\put(-25,130){\scriptsize positive energy}
	\put(-5,120){\scriptsize sector}
	\put(185,20){\scriptsize negative energy}
	\put(205,10){\scriptsize sector}
	\multiput(35,75)(0,10){7}{\line(1,0){15}}
	\multiput(150,5)(0,10){7}{\line(1,0){15}}
	\multiput(0,70)(10,0){9}{\line(1,0){5}}
	\multiput(115,70)(10,0){9}{\line(1,0){5}}
	\put(53,73){\scriptsize $0_+$}
	\put(53,83){\scriptsize $+1$}
	\put(53,93){\scriptsize $+2$}
	\put(53,103){\scriptsize $+3$}
	\put(53,113){\scriptsize $+4$}
	\put(53,123){\scriptsize $+5$}
	\put(53,133){\scriptsize $+6$}
	\put(168,63){\scriptsize $0_-$}
	\put(168,53){\scriptsize $-1$}
	\put(168,43){\scriptsize $-2$}
	\put(168,33){\scriptsize $-3$}
	\put(168,23){\scriptsize $-4$}
	\put(168,13){\scriptsize $-5$}
	\put(168,3){\scriptsize $-6$}
	\put(205,67){wall of $0$ (zero)}
	\end{picture}
	\end{center}
	\caption[tower of states]{Physical states in the two sectors}
	\label{tower}
\end{figure}

To avoid the misconceptions that the positive and negative energy sectors may simultaneously coexist and that there is no distinction between them, we depict them in Fig.~\ref{tower}.

Some explanation is in order. Figure~\ref{tower} displays towers that consist of states in the positive and negative energy particle states. In the positive energy particle states, transitions upward and downward occur through the actions of $a_+^{\dagger}(\vec{k})$ and $a_+(\vec{k})$ written by $A_+(\vec{k})$ respectively, while in the negative energy particle states, $a_-^{\dagger}(\vec{k})$ and $a_-(\vec{k})$ written by $A_-(\vec{k})$ produce transitions upward and downward, respectively. There exist positive and negative energy sectors in each energy particle states, as is understood from the method of the construction of the double harmonic oscillator. However, in the positive energy particle states depicted in Fig.~\ref{tower}, only the positive energy sector has physical meaning. Similarly, in the negative energy particle states, only the negative energy sector is realized in our real world. The reason is that there exists a wall of the classical number ``0 (zero)", or, in other word, a barrier of $e^{-\frac{1}{2}\! \int \! \frac{d^3\vec{k}}{(2\pi )^3} \left\{ a_-(\vec{k})+a_-^{\dagger}(\vec{k})\right\}^2}$, which cannot be penetrated with the usual polynomial interaction of field theories. Consequently, we can conclude that the total vacuum of the complex scalar boson system is given by the direct product (84) of the empty positive energy sector vacuum and the modified negative energy sector vacuum, so that $a_-^{\dagger}(\vec{k})||0_-\rangle =0$, i.e. to obey (86). The negative energy sector vacuum is a state in which an infinite number of negative energy bosons, in a sense, exist, as described by Eq.(93), and furthermore, by acting with $a_-(\vec{k})$, we can annihilate an infinite number of negative energy particles. Thus, there exist all numbers (including infinity) of excited states. In fact, in the boson theory, the relation (93), which relates the empty vacuum $||0_+\rangle$ and the sea vacuum $||0_-\rangle$, exponentiated operators is preferable and, indeed, needed.

Next, we study the possible values of the energy of the boson system for which the vacuum is represented by a sea, by constructing the energy-momentum operator. This operator is derived using Noether's theorem: 

\begin{align}
	P^{\mu} & =\int d^3\vec{x}\left\{ \frac{\partial \mathcal{L}}
	{\partial (\partial_{\mu}A_+(x))}\partial_0A_+(x)
	+\frac{\partial \mathcal{L}}{\partial (\partial_{\mu}A_-(x))}
	\partial_0A_-(x)-\eta^{\mu 0}\mathcal{L}\right\} 
	\nonumber \\
	& =\int d^3\vec{k}\> \frac{1}{2}k^{\mu}\left\{ a_+(\vec{k})
	a_+^{\dagger}(\vec{k})+a_+^{\dagger}(\vec{k})a_+(\vec{k})+a_-(\vec{k})
	a_-^{\dagger}(\vec{k})+a_-^{\dagger}(\vec{k})a_-(\vec{k})\right\}.
\end{align}

\noindent With the commutator (83), if we define 

\begin{align}
	P^{\mu}=\int d^3\vec{k}\> k^{\mu}\left\{ a_+^{\dagger}(\vec{k})
	a_+(\vec{k})+a_-(\vec{k})a_-^{\dagger}(\vec{k})
	+\int \frac{d^3\vec{x}}{(2\pi)^3}\> 1\right\},
\end{align}

\noindent we obtain a physically consistent particle picture. For instance, the energy-momentum of the vacuum $||0\rangle_{\text{boson}}=||0_+\rangle \otimes ||0_-\rangle$ is computed as 

\begin{align}
	P^{\mu}||0\rangle_{\text{boson}} & =\int d^3\vec{k}\> k^{\mu}\left\{ 
	a_+^{\dagger}(\vec{k})a_+(\vec{k})+a_-(\vec{k})a_-^{\dagger}(\vec{k})
	+\int \frac{d^3\vec{x}}{(2\pi)^3}\> 1\right\} 
	||0_+\rangle \otimes ||0_-\rangle \nonumber \\
	& =\int \frac{d^3\vec{k}d^3\vec{x}}{(2\pi)^3}\> k^{\mu}
	||0\rangle_{\text{boson}}.
\end{align}

\noindent It is worthwhile to note that the bosonic vacuum energy in (96) is exactly canceled by the fermionic one if we consider a supersymmetric hypermultiplet. 

The energy-momenta of the first excited states are calculated similarly: 

\begin{align}
	P^{\mu}a_+^{\dagger}(\vec{p})||0\rangle_{\text{boson}} & 
	=\int d^3\vec{k}\> k^{\mu}
	\left\{ a_+^{\dagger}(\vec{k})a_+(\vec{k})+a_-(\vec{k})a_-^{\dagger}
	(\vec{k})+\int \frac{d^3\vec{x}}{(2\pi)^3}\> 1\right\} 
	a_+^{\dagger}(\vec{p})||0\rangle_{\text{boson}} \nonumber \\
	& =\left\{p^{\mu}+\int \frac{d^3\vec{k}d^3\vec{x}}{(2\pi)^3}\> k^{\mu}
	\right\} a_+^{\dagger}(\vec{p})||0\rangle_{\text{boson}} , \\
	P^{\mu}a_-(\vec{p})||0\rangle_{\text{boson}} & =\int d^3\vec{k}\> 
	k^{\mu}\left\{ a_+^{\dagger}(\vec{k})a_+(\vec{k})+a_-(\vec{k})
	a_-^{\dagger}(\vec{k})+\int \frac{d^3\vec{x}}{(2\pi)^3}\> 1\right\} 
	a_-(\vec{p})||0\rangle_{\text{boson}} \nonumber \\
	& =\left\{p^{\mu}+\int \frac{d^3\vec{k}d^3\vec{x}}{(2\pi)^3}\> k^{\mu}
	\right\} a_-(\vec{p})||0\rangle_{\text{boson}}.
\end{align}

\noindent From these results, it is seen that the energy-momentum of the states excited under the action of $a_+^{\dagger}(\vec{p})$ and $a_-(\vec{p})$ are increased by an amount $p^{\mu}$. This implies the following: In the positive energy particle states, the interpretation is the usual one, in which one particle of energy-momentum $p^{\mu}$ is created on the empty vacuum. In the negative energy particle states, because in the vacuum state an infinite number of negative energy particles are filled, when $a_-(\vec{p})$ acts on this state, one negative energy particle is annihilated, creating a hole that is interpreted as an anti-particle, because this hole is considered to have energy-momentum $p^{\mu}$ relative to the surrounding particles with negative energy-momentum. All other higher states can be interpreted in the same manner. Therefore, defining the energy-momentum operator $P^{\mu}$ by Eq.(95), we obtain a physical particle picture that is consistent with that of the second quantization, in which all the excited states have positive energies. 

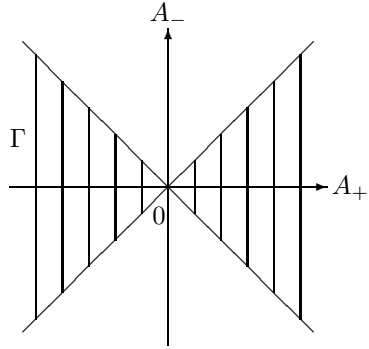
\begin{figure}[htb]
	\begin{center}
	\begin{picture}(120,140)
	\put(0,60){\vector(1,0){120}}
	\put(60,0){\vector(0,1){120}}
	\put(122,57.5){$A_+$}
	\put(53,123){$A_-$}
	\put(54,46){$0$}
	\put(0,75){$\Gamma$}
	\put(5,5){\line(1,1){110}}
	\put(5,115){\line(1,-1){110}}
	\put(70,50){\line(0,1){20}}
	\put(80,40){\line(0,1){40}}
	\put(90,30){\line(0,1){60}}
	\put(100,20){\line(0,1){80}}
	\put(110,10){\line(0,1){100}}
	\put(50,50){\line(0,1){20}}
	\put(40,40){\line(0,1){40}}
	\put(30,30){\line(0,1){60}}
	\put(20,20){\line(0,1){80}}
	\put(10,10){\line(0,1){100}}
	\end{picture}
	\end{center}
	\caption[region]{Region $\Gamma$ for which the inner product (99) 
	converges.}
	\label{Gamma}
\end{figure}

To end the present section, a comment about the inner product of the states in the second quantized theory is in order. If we write the $(n,m)$-th excited state as $||n,m\rangle \simeq \Phi_{n,m}[A_+,A_-]$ for the complex scalar field, where $n$ represents the number of the positive energy particles and $m$ that of negative ones, the inner product is defined by 

\begin{align}
	\langle n,m||n^{\prime},m^{\prime}\rangle =\int_{\Gamma} 
	\mathfrak{D}A_+ \mathfrak{D}A_- \> 
	\Phi_{n,m}^{\dagger}[A_+,A_-]\Phi_{n^{\prime},m^{\prime}}[A_+,A_-],
\end{align}

\noindent where on the right-hand side there appears a functional integration over the real scalar fields $A_+(\vec{k})$ and $A_-(\vec{k})$, which correspond to the coordinates $x$ and $y$, and $\Phi_{n,m}^{\dagger}$ means $\Phi_{n,m}^{\ast}$ as in the 2 dimensional harmonic oscillator system in Subsection 3.3. Here the integration region $\Gamma$ can be obtained as Fig. \ref{Gamma} by replacing $x$ by $A_+$ and $y$ by $A_-$ for region $\gamma$ in Fig. \ref{gamma}.

\section{One particle state with negative energy: \\The solution of the Klein-Gordon equation}

\vspace{0.5cm}

\indent In the previous sections, we have focused on the problem of negative energy particles in terms of the 2nd quantization of the scalar field theory. It seemed our formulation works very well. However, we must answer the remaining question as for the single particle state of boson. That is, as is well-known, the Klein-Gordon equation has the following difficulty: The Lorentz-invariant norm square, i.e. the density of probability, 

\begin{align}
	P_b(t)=\int d^3\vec{x}\> A^{\dagger}(x)\> 
	i\overleftrightarrow{\partial_0}A(x)=
	\int d^3\vec{k}\> \left\{ a_+^{\dagger}(\vec{k})a_+(\vec{k})
	-a_-(\vec{k})a_-^{\dagger}(\vec{k})\right\} ,
\end{align}

\noindent is not positive definite for the negative energy solution, since the energy $E=i\partial_0$ is negative. Is it true also in our formulation? Our answer is No{\bf !} The inner product for the one particle state with negative energy is positive definite in the same way as that for the Dirac particle, 

\begin{align*}
	P_f(t)=\int d^3\vec{x}\> \bar{\psi}(x)\> \gamma^0\psi (x).
\end{align*}

\begin{figure}[ht]
	\begin{center}
	\begin{picture}(200,180)
	\put(35,0){\line(0,1){140}}
	\put(50,0){\line(0,1){140}}
	\put(150,0){\line(0,1){140}}
	\put(165,0){\line(0,1){140}}
	\put(93,67){\Large $\otimes$}
	\put(40,-10){\Large $\vdots$}
	\put(40,140){\Large $\vdots$}
	\put(155,-10){\Large $\vdots$}
	\put(155,140){\Large $\vdots$}
	\put(15,170){\scriptsize potisive energy}
	\put(17,160){\scriptsize particle states}
	\put(130,170){\scriptsize negative energy}
	\put(132,160){\scriptsize particle states}
	\put(-25,130){\scriptsize positive energy}
	\put(-5,120){\scriptsize sector}
	\put(175,20){\scriptsize negative energy}
	\put(195,10){\scriptsize sector}
	\put(90,130){\scriptsize positive energy}
	\put(110,120){\scriptsize sector}
	\multiput(35,75)(0,10){7}{\multiput(0,0)(2,0){8}{\line(1,0){1}}}
	\multiput(150,5)(0,10){7}{\multiput(0,0)(2,0){8}{\line(1,0){1}}}
	\multiput(150,75)(0,10){7}{\multiput(0,0)(2,0){8}{\line(1,0){1}}}
	\put(150,85){\thicklines \line(1,0){15}}
	\put(150,85.5){\thicklines \line(1,0){15}}
	\put(198,118){\vector(-1,-1){30}}
	\put(198,118){\framebox(100,20)[t]{one particle state}}
	\put(198,118){\framebox(100,20)[b]{with negative energy}}
	\multiput(0,70)(10,0){9}{\line(1,0){5}}
	\multiput(115,70)(10,0){9}{\line(1,0){5}}
	\put(205,67){wall of $0$ (zero)}
	\end{picture}
	\end{center}
	\caption[1 particle state]{The solution of the Klein-Gordon equation}
	\label{one}
\end{figure}
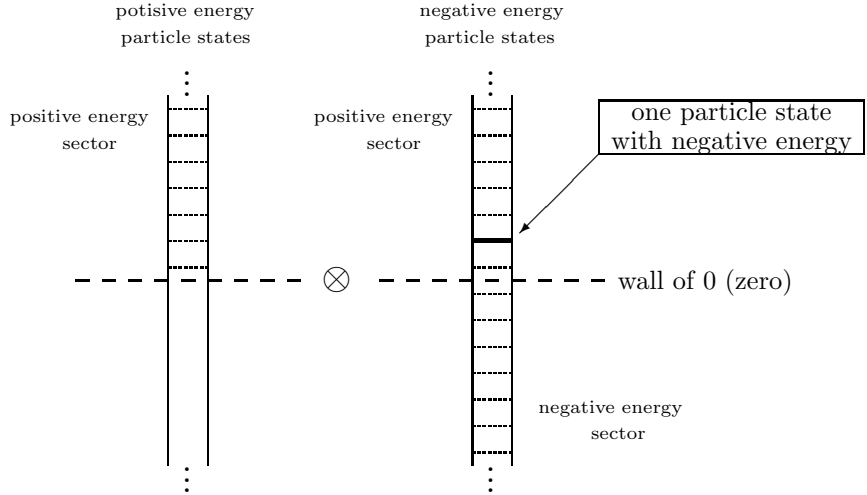

To explain why this is the case, let us certify the one particle state with negative energy in terms of the 2nd quantized theory, but in the harmonic oscillator language as Section 3 for simplicity. As we have seen, the vacuum $|0_-\rangle$ for the system of the negative energy particles is not empty and, in fact, is filled with negative energy particles, so the first excited state $a_-|0_+\rangle \otimes |0_-\rangle$ is not one particle state with negative energy but one hole state with positive one. Therefore the negative energy solution of the Klein-Gordon equation is represented as $a_-^{\dagger}|0_+\rangle \otimes |0_+\rangle $, where $a_-^{\dagger}$ acts on the latter $|0_+\rangle$, since $a_-^{\dagger}$ creates a negative energy particle and the latter $|0_+\rangle \simeq e^{-\! \frac{1}{2} A_-^2}$ is the empty vacuum in the negative energy particle states. However, recalling the representation of the double harmonic oscillator in Subsection 3.2, we have to make the inverse tranformation  of Eq.(40), 

\begin{align}
	a_-=i\otimes \frac{1}{\sqrt{2}}\left( A_-+\frac{d}{dA_-}\right) & \to 
	1\otimes \frac{1}{\sqrt{2}}\left( A_--\frac{d}{dA_-}\right)
	\equiv \check{a}_-^{\dagger}=-i\otimes a_-^{\dagger}, \nonumber \\
	a_-^{\dagger}=i\otimes \frac{1}{\sqrt{2}}\left( A_--\frac{d}{dA_-}
	\right) & \to 1\otimes \frac{1}{\sqrt{2}}\left( A_-+\frac{d}{dA_-}
	\right)\equiv \check{a}_-=-i\otimes a_-, \\
	& \nonumber \\
	[a_-,a_-^{\dagger}]=-1 & \to [\check{a}_-,\check{a}_-^{\dagger}]=+1, 
\end{align}

\noindent since this one particle state belongs to the positive energy sector in the negative energy particle states. Therefore,to tell the truth, it should be written as 

\begin{align}
	\check{a}_-^{\dagger}|0_+\rangle \otimes |0_+\rangle \equiv 
	|KG_-\rangle , 
\end{align}

\noindent which is depicted in Fig. \ref{one} by bold line above the ``wall of 0".

As seen in Eqs.(101) and (102), the positive energy sector in the negative energy particle states is completely similar to that in the positive ones as the Hilbert space, so all the states in this sector has the positive definite norms. For example, 

\begin{align}
	\langle KG_-|KG_-\rangle & 
	=\langle 0_+| \otimes \langle 0_+|\check{a}_-
	\check{a}_-^{\dagger}|0_+\rangle \otimes |0_+\rangle \nonumber \\
	& =\langle 0_+| \otimes \langle 0_+|\big \{\check{a}_-^{\dagger}
	\check{a}_-+1\big \}|0_+\rangle \otimes |0_+\rangle  \nonumber \\
	& =+1.
\end{align}

\noindent Of course, the inner products in the function representation is defined just like as Eq.(60), the ordinary ones. However the energy-momentum operator (95) and the density of probability (100) is changed in its form according to the inverse transformation (101). They are deformed to 

\begin{align}
	& \check{P}^{\mu}=\int d^3\vec{k}\> k^{\mu}\left\{ 
	a_+^{\dagger}(\vec{k})a_+(\vec{k})-\check{a}_-^{\dagger}(\vec{k})
	\check{a}_-(\vec{k})+\int \frac{d^3\vec{x}}{(2\pi)^3}\> 1\right\}, \\
	& \check{P}_b(t)=\int d^3\vec{k}\> \left\{ a_+^{\dagger}(\vec{k})
	a_+(\vec{k})+\check{a}_-^{\dagger}(\vec{k})\check{a}_-(\vec{k})
	\right\}.
\end{align}

\noindent This result means that the negative energy solution of the Klein-Gordon equation has the positive definite norm, i.e. dendity of probability as 

\begin{align}
	\check{P}_b(t)\check{a}_-^{\dagger}(\vec{p})||0_+\rangle \otimes 
	||0_+\rangle =(+1)\check{a}_-^{\dagger}(\vec{p})||0_+\rangle \otimes 
	||0_+\rangle ,
\end{align}

\noindent in our formalism. In the same way, all the higher excited states has the positive definite norms. Added to this, we can confirm that the operator $a_-=i\otimes \check{a}_-$ certainly annihilate a negative energy boson. That is, from the following calculation, 

\begin{align}
	\check{P}^{\mu}\check{a}_-^{\dagger}(\vec{p})||0_+\rangle \otimes 
	||0_+\rangle =\left\{-p^{\mu}+\int d^3\vec{k}\>k^{\mu}\> 
	\int \frac{d^3\vec{x}}{(2\pi)^3}\> 1\right\}
	\check{a}_-^{\dagger}(\vec{p})||0_+\rangle \otimes ||0_+\rangle ,
\end{align}

\noindent the operator $a_-^{\dagger}(\vec{p})=i\otimes \check{a}_-^{\dagger}(\vec{p})$ creates the negative energy boson which has the energy-momentum $(-p^{\mu})$, then we obtain the result that $a_-=i\otimes \check{a}_-$ is the annihilation operator of the negative energy boson also in this way.

\section{Conclusion and future perspectives}

\vspace{0.5cm}

In this paper, we have proposed the idea that the boson vacuum forms a sea, like the Dirac sea for fermions, in which all the negative energy states are filled. This was done by introducing a double harmonic oscillator, which stems from an extension of the concept of the wave function. Furthermore, analogous to the Dirac sea where due to the exclusion principle each negative energy state is filled with one fermion, in the boson case we also discussed a modification of the vacuum state so that one could imagine two types of different vacuum fillings for all the momenta. In the picture of the double harmonic oscillator, we imagine the analogy of filling the Dirac sea for fermions to be to replace for each negative single boson orbit the state $|0_+\rangle$ by $|0_-\rangle$, an operation that in~\cite{nn,nn2} were described as taking a boson away from the beginning vacuum. Then the usual interpretation of an anti-particle, as a hole in the negative energy sea, turns out to be applicable not only for the case of fermions but also for that of bosons. Thus, we have proposed a way of resolving the long-standing problem in field theory that the bosons cannot be treated analogously to the old Dirac sea treatment of the fermions and in the Klein-Gordon theory that the negative energy state has the indefinite norm. Our presentation relies on the introduction of the double harmonic oscillator, but that is really just to make it concrete. What is really needed is that we formally extrapolate to have negative numbers of particles in the single particle states, precisely what is described by our ``double harmonic oscillator", which were extended to have negative numbers of excitation quanta. Supersymmetry also plays a substantial role in the sense that it provides us with a guideline for how to develop the method. In fact, our method is physically very natural when we consider supersymmetry, which, in some sense, treats bosons and fermions on an equal footing. 

Our picture of analogy between fermion and boson sea description is summarized by Table 2.

\begin{center}
	\begin{tabular}{|c|c|c|} \hline
	& \multicolumn{2}{|c|}{\textbf{\large Fermions}} \\
	& \small{positive energy particle states} & 
	\small{negative energy particle states} \\
	& $E>0$ & $E<0$ \\ \hline
	& & \\
	empty $||\tilde{0}_+\rangle$ & true & not realized in nature \\
	& & \\ \hline
	& & \\
	filled $||\tilde{0}_-\rangle$ & not realized in nature & true \\
	& & \\ \hline
	\multicolumn{3}{c}{} \\ \hline
	& \multicolumn{2}{|c|}{\textbf{\large Bosons}} \\
	& \small{positive energy particle states} & 
	\small{negative energy particle states} \\
	& $E>0$ & $E<0$ \\ \hline
	& & \\
	empty $||0_+\rangle$ & true & not realized in nature \\
	& & {\scriptsize (include solution of Klein-Gordon eq.)}\\ \hline
	& & \\
	filled $||0_-\rangle$ & not realized in nature & true \\
	& & \\ \hline
	\multicolumn{3}{c}{} \\
	\multicolumn{3}{c}{Table 2: Analogy between fermion and boson sea 
	description} \\
	\end{tabular}
\end{center}

Our method for constructing the boson sea is expected to have a wide range of applications in quantum physics. First of all, string theories and string field theories have been successfully quantized using the light-cone quantization method~\cite{kaku}. However, for string field theories there are no satisfactory theories at the present time. Our present task is to clarify why only the light-cone quantization can be carried out. In other words, we would like to determine the essential points that make covariant quantization so difficult. The first quantization of the string world sheet coordinates $X^{\mu}(\tau ,\sigma)$ is performed through 

\begin{align*}
	& [X^{\mu}(\tau ,\sigma ),\Pi^{\nu}(\tau ,\sigma^{\prime})]
	=i\eta^{\mu \nu}\delta (\sigma -\sigma^{\prime}), \\
	& \quad \eta^{\mu \nu}=diag(-1,+1,\cdots ,+1). \nonumber
\end{align*}

\noindent In the light-cone quantization the negative energy states that stems from $(\mu ,\nu )\! =\! (0,0)$ component of the target space metric $\eta^{\mu \nu}$ disappear. This mechanism precisely coincides with the commutation relation (83) for the complex scalar field theory. In view of the latter theory usage of the commutator with $-1$ on the right hand side (83) means to consider the negative energy sector, i.e. boson sea. In attacking these problems, our method may be useful, because we believe that perhaps the reason why only the light-cone quantization is successful is that it removes all the negative energy states. Therefore, to carry out the covariant quantization, we should treat the negative energy states in a sophisticated manner. Our method may be useful for this purpose. This is one of the main motivations to develop our method of treating the old problem of quantizing bosons by appropriately including negative energy states, in analogy to the Dirac's method for fermions. Also in analogy with the intuitive understanding and derivation of the chiral anomaly of the massless fermion as pair creation from the Dirac sea~\cite{nn}, we may expect to obtain a new insight about boson anomaly such as the conformal anomaly. Furthermore there is a possibility that the boson propagator may be modified by the effect of the boson sea. In fact the integration region $\Gamma$ in Fig. \ref{Gamma} is nontrivial due to the effect of the boson sea under the definition of the norm (99) in the Hilbert space.

\vspace{2cm}

\noindent \underline{ Acknowledgement }

\vspace{0.25cm}

This work is supported by Grants-in-Aid for Scientific Research on Priority Areas, Number of Areas 763, ``Dynamics of strings and Fields", from the Ministry of Education of Culture, Sports, Science and Technology, Japan.

\vspace{0.5cm}


\end{document}